\documentclass[twocolumn,twocolappendix]{aastex701}

\begin{document}

\title{Discovery of a Candidate 2 keV Cyclotron Resonance Scattering Feature in the HLX NGC~3583~X-1}

\author[orcid=0000-0002-3430-9837,gname=Kiran,sname=Jayasurya]{Kiran M. Jayasurya}
\affiliation{Astronomy and Astrophysics, Raman Research Institute, C. V. Raman Avenue, Sadashivanagar, Bangalore 560080, India}
\affiliation{Space Astronomy Group, ISITE Campus, U. R. Rao Satellite Centre, ISRO, Bengaluru 560037, India}
\email[show]{mkiran.jayasurya@gmail.com}

\author[orcid=0009-0000-2039-4340,gname=Aman,sname=Upadhyay]{Aman Upadhyay}
\affiliation{Astronomy and Astrophysics, Raman Research Institute, C. V. Raman Avenue, Sadashivanagar, Bangalore 560080, India}
\email{aman.upadhyay@rrimail.rri.res.in}

\author[orcid=0000-0003-1703-8796,gname=Vikram,sname=Rana]{Vikram Rana}
\affiliation{Astronomy and Astrophysics, Raman Research Institute, C. V. Raman Avenue, Sadashivanagar, Bangalore 560080, India}
\email{vrana@rri.res.in}

\begin{abstract}
We present a broadband X-ray study of the transient hyperluminous X-ray source (HLX), 2SXPS~J111416.1+481833, in the galaxy NGC~3583, using archival \textit{XMM-Newton}, \textit{NuSTAR}, \textit{Chandra} data, and long-term \textit{Swift}/XRT monitoring. The source episodically enters the hyperluminous regime with X-ray luminosities $L_{X}>10^{41}$~erg~s$^{-1}$ and drops by a factor of $>45$ from its peak into a deep low state.  We detect a clear spectral cutoff at $\sim5$--6~keV in the broadband spectra, which are well modeled by a soft thermal component combined with optically thick thermal Comptonization or an inner advection-dominated disk. In the \textit{XMM-Newton} spectra, we detect a statistically significant ($ \gtrsim 3.9 \sigma$) absorption line centered at $E_{\rm line} \approx 1.97 \pm 0.04$ keV with a width of $\sigma_{\rm line} \approx 74 \pm 40$~eV. We primarily interpret the line as a candidate proton Cyclotron Resonance Scattering Feature~(CRSF), implying a local magnetic field strength of $B \sim 4 \times 10^{14}$~G. Alternative interpretations, such as an origin in an ionized outflow, were explored and found to be less likely. We do not detect coherent X-ray pulsations, placing 90\% confidence upper limits on the pulsed fraction of 19.3\% in the 0.3--10~keV band and 36.3\% in the 3--15~keV band. The combination of extreme luminosity, a hard spectral state, and the detection of a candidate cyclotron line provides strong evidence for a highly magnetized neutron star accretor.
\end{abstract}

\keywords{\uat{Ultraluminous X-ray sources}{2164} --- \uat{Neutron stars}{1108} --- \uat{X-ray binaries}{1811} --- \uat{X-ray sources}{1822} ---  \uat{X-ray astronomy}{1810} --- \uat{High Energy astrophysics}{739}}



\section{Introduction}
\label{sec:intro}

Extremely bright, off-nuclear, point-like X-ray sources with luminosities in the range $L_{\rm X} \sim 10^{39}$--$10^{41}$~erg~s$^{-1}$ are collectively known as ultraluminous X-ray sources (ULXs; see \citealt{feng2011,kaaret2017,king2023} for recent reviews). Such high luminosities initially motivated the interpretation that ULXs might host intermediate-mass black holes (IMBHs; $M \sim 10^{2}$--$10^{5}\,M_{\odot}$; \citealt{colbert1999,makishima2000,miller2004}). However, alternative scenarios such as beaming and super-Eddington accretion were also proposed \citep{king2001, begel2002, kording2002, ebisawa2003}.

In recent decades, broadband X-ray spectroscopy has revealed that many ULXs show a clear spectral curvature or cutoff at relatively low energies ($<10$~keV; \citealt{stobbart2006,gladstone2009}), which can be naturally explained by super-Eddington accretion onto stellar-mass compact objects such as black holes (BHs) or neutron stars (NSs). The discovery of coherent pulsations in the ULX M82~X-2 \citep{bachetti2014} and other pulsating ULXs (PULXs; \citealt{furst2016,israel2017,israel2017b,carpano2018,satya2019,castillo2020,ducci2025}) unambiguously demonstrated that neutron stars can power X-ray luminosities well beyond their classical Eddington limits.

A rare class of sources exhibiting X-ray luminosities $L_{\rm X} > 10^{41}$~erg~s$^{-1}$ has been categorized as hyperluminous X-ray sources (HLXs). The HLX population appears to consist of two types of objects. The first type, such as ESO~243-49~HLX-1 \citep{farrell2009,servi2011}, shows properties reminiscent of sub-Eddington accretion in massive black hole binaries. These are thought to be strong IMBH candidates \citep{sutton2012}. In contrast, the second type of HLX reaches hyperluminous levels episodically and otherwise resembles extreme ULXs, such as NGC~5907~ULX1 \citep{israel2017,furst2023}. These objects are thought to be super-Eddington accretors and may potentially represent the high-luminosity tail of the ULX population \citep{amato2025}. A notable recent addition to this category is NGC~470~HLX1 \citep{ghosh2025}.

At super-Eddington accretion rates, the inner accretion disk becomes thick and advection-dominated, resulting in the formation of a funnel-like structure along the polar axis. This geometry causes the radial temperature profile to deviate from the standard thin-disk relation ($T \propto r^{-0.75}$; \citealt{shakura1973}) to a flatter profile ($T \propto r^{-p}$ with $p<0.75$; \citealt{watarai2000,watarai2001}). 

In many ULXs, such as NGC~1313~X1, NGC~4395~ULX1, and M74~X-1, residuals near 1~keV cannot be adequately explained by continuum curvature alone \citep{middle2015r, ghosh2022, aman2025}.
Systematic studies of \textit{XMM-Newton} RGS spectra have revealed atomic lines, such as O~\textsc{viii}, Ne~\textsc{x}, and Fe~\textsc{xxv}, with blueshifts corresponding to outflow velocities of $\sim 0.1$--$0.3c$ \citep{pinto2017,kosec2018,kosec2021,bright2022,pinto2023}. These results provide strong evidence for large-scale, radiation-driven outflows or winds launched by the accretion disk, as predicted by models of super-Eddington accretion \citep{poutanen2007,ohsuga2007,ohsuga2011}.

Alternatively, absorption features have also been interpreted as cyclotron resonance scattering features (CRSFs; \citealt{truemper1978,meszaros1992}). CRSFs originate when X-ray photons are resonantly scattered at characteristic energies by charged particles in a strong magnetic field. The cyclotron line energy ($E_{\rm cyc}$) provides a relatively direct estimate of the magnetic field strength of the neutron star \citep{meszaros1992, staubert2019}. Electron CRSFs (e-CRSFs) are produced by electrons in magnetic fields of $B \sim 10^{11}$--$10^{13}$~G, typically at energies of $E_{\rm cyc,e} \sim 10$--$150$~keV, with a line-width-to-energy ratio of $\sigma/E_{\rm cyc,e} \sim 0.1$--$0.5$ \citep{coburn2002,staubert2019}. In contrast, proton CRSFs (p-CRSFs), due to the much larger proton mass, are expected to appear at lower energies of $E_{\rm cyc,p} \sim 0.5$--$5$~keV with narrower widths ($\sigma/E_{\rm cyc,p} < 0.1$; \citealt{ibra2002}), requiring magnetar-level magnetic fields ($B \sim 10^{14}$--$10^{15}$~G). 

Observationally, \cite{bright2018} reported the first proton CRSF in a ULX, namely M51~ULX8, at $E_{\rm cyc} \sim 4.5$~keV, implying a magnetar-strength field of $B \sim 10^{15}$~G. However, \cite{middle2019} ruled out a global magnetar field for the source through covariance spectroscopy. Recently, \cite{cruz2026} reported a 3.3~keV p-CRSF feature in NGC~4656~ULX-1, implying a local magnetic field strength of $(6-7)\times 10^{14}$~G.

The source 2SXPS~J111416.1+481833 (hereafter NGC~3583~X-1) in the spiral galaxy NGC~3583 is located at RA = 11$^{\rm h}$14$^{\rm m}$16.1$^{\rm s}$ and Dec. = +48$^{\circ}$18$\arcmin$33.6$\arcsec$. It was listed as a ULX candidate in the catalog\footnote{\href{https://vizier.cds.unistra.fr/viz-bin/VizieR-5?-ref=VIZ6979a176125f89\&-out.add=.\&-source=J/A+A/681/A16/tablea2\&recno=2084}{Tranin et al. 2024: \ Table A.2, \#2084.}} compiled by \cite{tranin24} and has been a target for pointed \textit{XMM-Newton} observations from as early as 2022.

In this paper, we present a detailed X-ray analysis of NGC~3583~X-1 using \textit{XMM-Newton}, \textit{NuSTAR}, and \textit{Chandra} data, and long-term \textit{Swift}/XRT monitoring. We describe the observations and data reduction in Section~\ref{sec:obs}, the timing analysis and pulsation search results in Section~\ref{sec:pulsations}, the spectral modeling, including the significant detection of an absorption feature, in Section~\ref{sec:spec}, and finally, discuss the implications for the nature of the underlying compact object in Section~\ref{sec:disc}. We adopt a distance\footnote{\href{http://atlas.obs-hp.fr/hyperleda/fG.cgi?n=a007&o=NGC3583}{HyperLeda Database: NGC3583 Distances}}  of 31.6 $\pm$ 4.5~Mpc to NGC~3583 throughout this work, which was derived using the method of sosie galaxies  \citep{terry2002}.

\section{Observations and Data Reduction} \label{sec:obs}

We used multi-epoch data of NGC~3583~X-1 obtained from deep X-ray observations (see Table~\ref{tab:obs_log_deep} for details) alongside long-term \textit{Swift}/XRT monitoring data.  

\begin{deluxetable*}{lccccc}
 \tablecaption{Log of the X-ray observations of NGC~3583~X-1 used in this work. }
\label{tab:obs_log_deep}
  \tablewidth{0pt}
  \tablehead{
    \colhead{Observatory} &
    \colhead{Instrument} &
    \colhead{ObsID} &
    \colhead{Start Date} &
    \colhead{Exposure} & \colhead{Source Counts} \\
    \colhead{} &
    \colhead{} &
    \colhead{} &
    \colhead{(YYYY-MM-DD)} &
    \colhead{(ks)} & \colhead{}
  }
  \startdata
  \textit{Chandra} & ACIS-S & 19381 & 2017-02-09 & 9.9 & 36 \\
  \hline
  \textit{XMM-Newton} & PN & 0884070201 & 2022-04-23 & 26.5 & 5492\\
  & MOS1 & & & 32.2 & 1815 \\
  & MOS2 & & & 32.3 & 1836 \\
  \hline
  \textit{XMM-Newton} & PN & 0931440101 & 2024-05-30 & 35.3 & 8378 \\
  & MOS1 & & & 43.3 & 2997 \\
  & MOS2 & & & 49.9 & 3386  \\
  \hline
  \textit{NuSTAR} & FPMA & 50901001002 & 2024-05-30 & 183.7 & 1195 \\
  & FPMB & & & 183.8 & 1185 \\
  \enddata
  \tablecomments{The exposure times indicate the effective duration after filtering for background flares and applying Good Time Intervals (GTIs). Source counts correspond to the net, background-subtracted photon counts extracted in the 0.3--7.0 keV band for \textit{Chandra}, the 0.3--10.0 keV band for \textit{XMM-Newton} instruments, and in the 3.0--15.0 keV band for the \textit{NuSTAR} FPMA and FPMB modules.}
\end{deluxetable*}

\subsection{Swift/XRT}

We utilized the online automated pipeline tool\footnote{\href{https://www.swift.ac.uk/user_objects/}{UKSSDC: Build Swift-XRT products}} provided by the UK Swift Science Data Centre (UKSSDC; \citealt{evans2009}) to generate science-ready products for all available historical \textit{Swift}/XRT monitoring data of NGC~3583~X-1 between November 2015 and May 2024. Those observations with an exposure time $<$700~s and a pointing offset greater than 5$\arcmin$ were excluded from further analysis.
\begin{figure*}
  \epsscale{1}
  \plotone{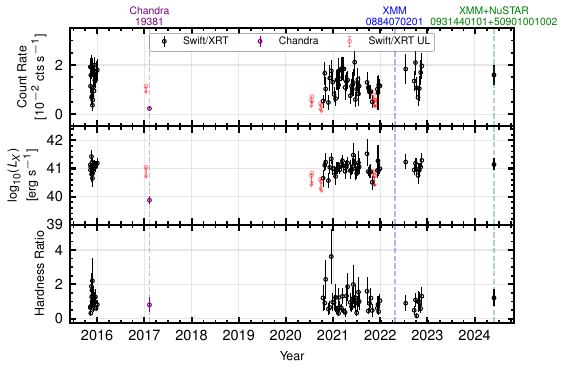}
  \caption{Long-term X-ray variability of NGC~3583~X-1. \textit{Top Panel}: Evolution of the 0.3--10 keV count rate. Black circles denote \textit{Swift}/XRT detections, red circles with downward arrows indicate $3\sigma$ upper limits for non-detections, and the purple circles denote the 2017 \textit{Chandra} observation (converted to equivalent \textit{Swift}/XRT rate). \textit{Middle Panel}: Evolution of the logarithm of the unabsorbed X-ray luminosity ($L_{\rm X}$) in the 0.3--10 keV band. \textit{Bottom Panel}: Evolution of the Hardness Ratio (ratio of the 1.5--10 keV count rate to the 0.3--1.5 keV count rate). 
  Vertical dashed lines mark the epochs of the deep X-ray observations analyzed in this work. For the count rate and HR panels, the error bars are 1$\sigma$ uncertainties from the \textit{Swift}/XRT data products and for the luminosity panel, error bars represent 90\% confidence intervals from the spectral fits. See text for more details.
  \label{fig:lteswift}}
\end{figure*}

Optimal background regions were determined dynamically by the pipeline  \citep{evans2007, evans2009}. To generate the light curve products, we used the time binning method, requiring at least 20 counts per bin and a minimum Signal-to-Noise Ratio (SNR) of 2.5. The light curve products provided count rates in the 0.3--10~keV, 0.3--1.5~keV (soft), 1.5--10~keV (hard) energy bands as well as the hardness ratio ($HR$), defined as the ratio of the hard-band count rate to the soft-band count rate. Source and background spectra as well as response files were generated for each observation. 

Due to the low photon statistics of the \textit{Swift}/XRT data, we grouped the spectra to a minimum of 1~count per bin and performed the spectral fitting using the $C$-statistic (\citealt{cash1979}; implemented as the $W$-statistic, \citealt{wachter1979}) in \texttt{XSPEC} (v12.15.1; \citealt{arnaud1996}), distributed within \texttt{HEASoft} v6.36 \citep{heasarc2014}. 

We modeled the spectra using an absorbed power-law, \texttt{TBabs*powerlaw}. Due to the low photon counts in the exposures, the neutral hydrogen column density ($N_{\rm H}$) and photon index ($\Gamma$) were often found to be strongly correlated or unconstrained. For consistency, we fixed $N_{\rm H} = 1.0 \times 10^{21}$~cm$^{-2}$ for all observations, corresponding to the best-fit column density determined from the high-quality \textit{XMM-Newton} observations. Unabsorbed luminosities in the 0.3--10~keV band were obtained using the \texttt{cglumin} model component in \texttt{XSPEC}. Observations with fewer than 5 source counts were treated as null detections. For these, we derived 3$\sigma$ upper limits by scaling the count rate uncertainties provided in the light curve products, using an average count-to-luminosity conversion factor.

The long-term temporal variation of the 0.3--10~keV count rate, the corresponding unabsorbed luminosity, and the hardness ratio ($HR$) are shown in Figure~\ref{fig:lteswift}. NGC~3583~X-1 is found to have an unabsorbed X-ray luminosity spanning $L_{\rm X}= 3.2^{+2.3}_{-1.5} \times 10^{40}$~erg~s$^{-1}$ (\textit{Swift}/XRT ObsID: 00034153062) to $L_{\rm X}=3.4^{+8.0}_{-2.1} \times 10^{41}$~erg~s$^{-1}$ (\textit{Swift}/XRT ObsID: 00034153058). This extreme luminosity firmly classifies the source as a transient HLX, exhibiting behavior similar to NGC~5907~ULX1 \citep{israel2017,furst2023}.

\subsection{Chandra}

\textit{Chandra} observed NGC~3583~X-1 on 2017 February 09 (ObsID 19381) using the Advanced CCD Imaging Spectrometer-S (ACIS-S) instrument for an effective exposure of $\sim$9.9 ks. We used the \textit{Chandra} Interactive Analysis of Observations (CIAO v4.17; \citealt{fruscione2006}) with the CALDB v4.12.2 for the data reduction.

We ran the \texttt{chandra\_repro} task to reprocess the data, apply the latest calibration, and create level-2 event files. The source region was defined as a circular region of $2''$ centered on the source coordinates, while the background was extracted from a $10\arcsec$ circular region in a source-free region on the same CCD close to the source. 

The spectra and the response files were generated using the \texttt{specextract} task. The spectra were grouped to a minimum of 1 count per bin, and the spectral fitting was done in the 0.3--7.0 keV energy range using $C$-statistic due to low degrees of freedom.

An absorbed power-law model (\texttt{TBabs*powerlaw}) was used to fit the spectrum, resulting in a $C$-stat/d.o.f ($C$-statistic/degrees of freedom) of 31.4/32. Since $N_{\rm H}$ was unconstrained, it was fixed at $ 1 \times 10^{21}$~cm$^{-2}$ to maintain consistency with the \textit{Swift}/XRT analysis. The source was found to be in a low state compared to the \textit{Swift}/XRT monitoring with a model-predicted unabsorbed luminosity of $L_{\rm X} = 7.4^{+2.9}_{-1.9} \times 10^{39}$~erg~s$^{-1}$ (in the 0.3--10 keV energy band) and a steep power-law photon index ($\Gamma = 2.22^{+0.66}_{-0.62}$).

The derived count rate, luminosity, and hardness ratio from this observation are shown in Figure~\ref{fig:lteswift}. The \textit{Chandra} count rate was converted to an equivalent \textit{Swift}/XRT rate using the WebPIMMS tool \citep{mukai1993} for consistency.

\subsection{XMM-Newton}

\textit{XMM-Newton} observed NGC~3583~X-1 in two epochs, once in 2022 (ObsID: 0884070201) and again in 2024 (ObsID: 0931440101; see Table \ref{tab:obs_log_deep} for details). The European Photon Imaging Camera (EPIC) instruments (EPIC-pn, EPIC-MOS1, and EPIC-MOS2; \citealt{struder2001, turner2001}) were operated in \texttt{PrimeFullWindow} mode. The EPIC-pn camera used the \texttt{Thin1} filter during the 2022 observation and the \texttt{Medium} filter during the 2024 observation. The EPIC-MOS cameras were configured with the \texttt{Medium} filter during both epochs. For the 2022 epoch, the MOS data for both instruments consisted of single primary science exposures. For the 2024 MOS1 data, we used only the third exposure (U004), which corresponds to the primary science observation. The first exposure was too short, and the second exposure was in calibration mode (using the \texttt{CalClosed} filter).

We carried out the data reduction using the \textit{XMM-Newton} Science Analysis System (SAS v22.1.0; \citealt{gabriel2004}) software package and the latest Current Calibration Files (CCF). We used the \texttt{epproc} and \texttt{emproc} tasks to generate calibrated event lists. Periods of high background flaring were identified and removed\footnote{For the 2022 observation, the flare filtering removed 7.4 ks of exposure from PN, 1.4 ks from MOS1, and 1.3 ks from MOS2. For the 2024 observation, the corresponding removed exposures were 35.2 ks, 6.9 ks, and 19.7 ks, respectively.} using the \texttt{espfilt} task. The clean flare filtered exposure times are shown in Table \ref{tab:obs_log_deep}.

Source events were extracted from a circular region of radius $30\arcsec$ centered on the source position. Background events were extracted from a nearby source-free circular region of radius $60\arcsec$ on the same CCD chip. Standard event selections were applied (\texttt{PATTERN$\leq4$} for pn and \texttt{PATTERN$\leq12$} for MOS), together with the conservative filtering criterion \texttt{FLAG==0}. The flare-filtered event lists were examined for pile-up using the \texttt{epatplot} task, and no significant pile-up was detected in any of the EPIC instruments.

The source spectra and light curves were extracted using the \texttt{evselect} task. The response files were generated using the \texttt{rmfgen} and \texttt{arfgen} tasks. For the timing analysis, we used the \texttt{barycen} task to correct for photon arrival times to the solar system barycenter using DE405 ephemeris. Background-subtracted light curves in the 0.3--10.0~keV band were generated using the \texttt{epiclccorr} task.

We followed the standard procedure outlined in the \textit{XMM-Newton} data analysis threads\footnote{\url{https://www.cosmos.esa.int/web/xmm-newton/sas-thread-rgs}} to reduce the \textit{XMM-Newton}/RGS data and extract the first- and second-order spectra. In both epochs, we obtained fewer than 450 net source counts per RGS instrument, and the spectra were found to be overwhelmingly dominated by the background. Notably, $\gtrsim 5000$ counts are required in the RGS spectra of ULXs to detect the strongest lines \citep{kosec2021,pinto2026}. Consequently, the RGS data were not used for high-resolution spectral modeling.

\subsection{NuSTAR}

The Nuclear Spectroscopic Telescope Array (\textit{NuSTAR}; \citealt{harrison2013}) observed NGC~3583~X-1 on 2024 May 30 (ObsID 50901001002) simultaneously with \textit{XMM-Newton}. The data were reduced using the \texttt{nupipeline} routine of NuSTARDAS (v2.1.2), distributed with HEASoft v6.36, and CALDB version 20251215. To filter out periods of high background activity caused by the South Atlantic Anomaly (SAA), we used the \texttt{saacalc=3}, \texttt{saamode=optimized}, and \texttt{tentacle=yes} flags. Since the default filtering criteria may not completely remove solar flaring activity, we screened the data by extracting background light curves in the 3--79 keV band in 100 s bins and excluded intervals where the count rate exceeded 0.2 counts~s$^{-1}$, just above the quiescent background level. After applying these filtering steps, the effective exposure was $\sim$184~ks in both Focal Plane Modules A (FPMA) and B (FPMB).

Since the background dominated beyond 15 keV, we limited our analysis of \textit{NuSTAR} data to the 3.0--15.0 keV energy band. The source products were extracted using the \texttt{nuproducts} task from a circular region with a radius of 30$\arcsec$ centered on the source position. Background spectra and light curves were extracted from a larger, source-free circular region of 60$\arcsec$ radius on the same detector chip. Photon arrival times were corrected to the solar system barycenter using the \texttt{barycorr=yes} option in \texttt{nuproducts} and the observation-specific orbit files. 

\section{Timing Analysis} 
\label{sec:pulsations}

\begin{figure}[ht!]
\centering
\begin{tabular}{@{}c@{}}
    \includegraphics[width=0.47\textwidth]{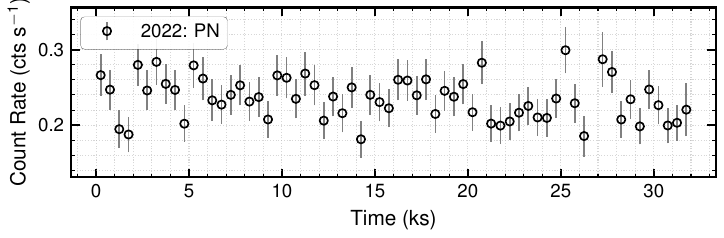} \\
    \includegraphics[width=0.47\textwidth]{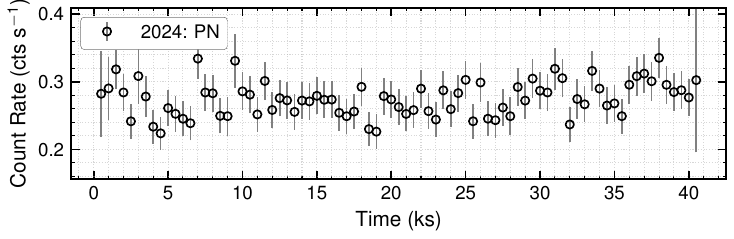} \\
    \includegraphics[width=0.47\textwidth]{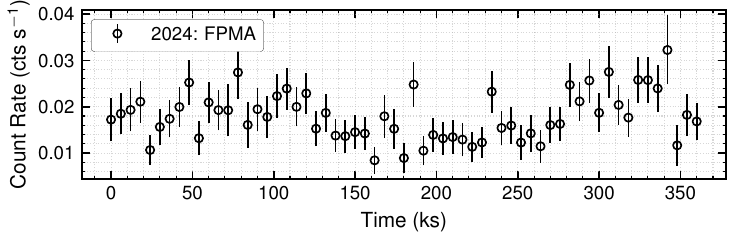}
    
\end{tabular}
\caption{Barycentric-corrected and background-subtracted light curves for NGC~3583~X-1. The top and middle panels show the 0.3--10 keV EPIC-pn data for the 2022 and 2024 epochs, respectively, binned at 500 s. The bottom panel displays the 3--15 keV \textit{NuSTAR} FPMA data binned at 6~ks. The horizontal axis represents time in kiloseconds (ks), and the vertical axis shows the count rate in counts~s$^{-1}$.}
\label{fig:lcxmm}
\end{figure}

The 0.3--10 keV \textit{XMM-Newton} EPIC-pn light curves binned at 500~s for epochs 2022 and 2024 are shown in the top and middle panels of Figure \ref{fig:lcxmm}, respectively. We also show the \textit{NuSTAR}~FPMA light curve in the 3--15 keV band for the source binned at 6~ks in the bottom panel of Figure \ref{fig:lcxmm}. 

To check for quasi-periodic oscillations (QPOs), we generated power density spectra (PDS) from the background-subtracted \textit{XMM-Newton} EPIC-pn and \textit{NuSTAR} light curves of both epochs using the \texttt{powspec} task within \texttt{HEASoft}. The features in the PDS were consistent with Poisson white noise, and no significant QPOs were detected.

\subsection{Pulsation Search}

To check for the presence of coherent pulsations, we extracted barycenter-corrected event lists corresponding to the source region from the \textit{XMM-Newton} EPIC-pn and \textit{NuSTAR} data. A blind pulsation search was carried out using the \texttt{HENDRICS} software package \citep{bachetti2018}, which is based on \texttt{Stingray} \citep{hupp2019}. We explored the frequency range of $f=0.01$--$6.5$~Hz ($P \approx 0.15$--100~s), encompassing the period range observed in known PULXs ($P\approx0.4$--30~s; \citealt{ducci2025}). We used the 0.3--10~keV band for the \textit{XMM-Newton} data and the 3--15~keV band for the \textit{NuSTAR} data.

We first performed an accelerated Fourier search based on \citet{Ransom2002} using \texttt{HENaccelsearch}, which is designed to detect sinusoidal modulations while accounting for frequency derivatives ($\dot{f}$) induced by orbital motion. For both epochs and all instruments, none of the candidate frequencies suggested by the algorithm produced a Fourier power above the significance threshold and all were consistent with Poisson noise. 

We also carried out a more sensitive pulsation search using the $Z^{2}_{2}$ statistic via the \texttt{HENzsearch} task. No significant coherent pulsations were detected in the explored frequency range. To account for the possibility of transient pulsations, we repeated the $Z^{2}_{2}$ searches on shorter data segments of 20~ks, 30~ks, and 60~ks. However, these segment-based analyses also yielded no significant detections.

Following the methodology of \cite{cruz2026h,cruz2026}, we computed 90\% confidence upper limits on the pulsed fractions through 10,000 Monte Carlo simulations using the \texttt{HENz2vspf} task. The resulting upper limits were found to be $\approx23.5\%$ and $\approx19.3\%$ in the \textit{XMM-Newton} 0.3--10~keV data from the 2022 and 2024 epochs, respectively. The corresponding upper limit for the combined \textit{NuSTAR} (FPMA+FPMB) data from 2024 was $\approx 36.3\%$ in the 3--15~keV energy band. These values are comparable to those reported for other ULXs without detected coherent pulsations \citep{earn2022,earn2024,cruz2026h, cruz2026}.

\section{Spectral Analysis}
\label{sec:spec}

We performed a detailed spectral analysis of NGC~3583~X-1 in \texttt{XSPEC} across two epochs using \textit{XMM-Newton} data from 2022 and joint \textit{XMM-Newton}+\textit{NuSTAR} data from 2024. To account for the absorption along the line-of-sight, we used the \texttt{TBabs} component, adopting abundances from \citet{wilms2000} and photoelectric cross-sections from \citet{verner1996}. For consistency, all spectra were grouped to a minimum of 1 count per bin using the \texttt{grppha} task and fitted with the $C$-statistic similar to \cite{bright2022}. 

A cross-normalization constant was included in all joint fits to account for calibration differences between the \textit{XMM-Newton} EPIC and \textit{NuSTAR} FPM detectors. The constant for EPIC-pn was fixed to unity, while the constants of the other instruments were allowed to vary. The resulting constant values remained within the expected $\sim 5-10$\% calibration uncertainties. All errors are quoted at the 90\% confidence level unless explicitly stated otherwise.

\subsection{2022: \textit{XMM-Newton} Observation}

The \textit{XMM-Newton} spectra were analyzed in the 0.3--8.0 keV energy band beyond which the background dominated. The EPIC spectra were initially modeled with a simple absorbed power-law (\texttt{TBabs*powerlaw}; Model M1), resulting in a photon index of $\Gamma = 1.9 \pm 0.05 $ and a $C$-stat/d.o.f of $ 1939.6/2400$. While statistically acceptable, we replaced the power-law with a cutoff power-law component (\texttt{TBabs*cutoffpl}; Model M2) to check for the presence of spectral curvature, as seen in ULXs \citep{stobbart2006,gladstone2009}. However, the fit statistic did not improve, and the high-energy cutoff ($E_{\rm cut}$) was unconstrained ($>100$ keV).

We next fitted the spectra with a physically motivated Comptonization model (\texttt{TBabs*compTT}; \citealt{titar1994}). However, the resulting $N_{\rm H}$ was unconstrained, indicating the need for an additional soft component. We also performed the fitting with an advection-dominated inner disk model (\texttt{TBabs*diskpbb}; \citealt{mineshige1994,watarai2000,watarai2001}), which resulted in a $C$-stat/d.o.f of 1943.6/2399 with a radial temperature profile exponent, $p \approx 0.53$, indicative of super-Eddington accretion, and an inner disk temperature of $T_{\rm in} \approx 3.2$ keV.

Since a single-component model such as \texttt{diskpbb} or \texttt{cutoffpl} can struggle to fit both the soft and hard continua simultaneously, we tested whether the continuum preferred an additional soft component in the form of a standard, geometrically thin, optically thick accretion disk (\texttt{diskbb}). The addition of the disk component to the \texttt{TBabs*cutoffpl} and \texttt{TBabs*diskpbb} models significantly improved the fit, with $\Delta C < -12.8$ for 2 additional degrees of freedom, in both cases. Moreover, the addition of the disk component to \texttt{TBabs*compTT} resulted in the $N_{\rm H}$ becoming constrained and consistent with the other models within uncertainties, yielding a $C$-stat/d.o.f $=1924.8/2397$.

We therefore considered standard two-component models commonly used to describe ULX continua, namely, \texttt{TBabs*(diskbb+cutoffpl)} (Model M3), \texttt{TBabs*(diskbb+diskpbb)} (Model M4), and \texttt{TBabs*(diskbb+compTT)} model (Model M5). For a BH accretor, the \texttt{cutoffpl} or \texttt{compTT} represents hard emission from a Comptonized corona, whereas for a NS accretor, they represent emission from an accretion column \citep{mushtu2015}. In both scenarios, the \texttt{diskbb} component likely represents emission from a cooler outer disk or the photosphere of an optically thick wind, and the \texttt{diskpbb} component represents emission from an inner, advection-dominated, hot disk forming a puffed-up funnel-like structure due to the super-Eddington accretion.

\begin{deluxetable*}{llccccc}
  \tabletypesize{\scriptsize}
  \setlength{\tabcolsep}{4pt}
  \tablecaption{Best-fit spectral parameters for the 2022 \textit{XMM-Newton} observation (ObsID 0884070201). 
  \label{tab:spectral_fits_2022}}
  \tablehead{
    \colhead{Component} & \colhead{Parameter} & \multicolumn{5}{c}{Model} \\
    \hline
    \colhead{} & \colhead{} & \colhead{M1} & \colhead{M2} & \colhead{M3} & \colhead{M4} & \colhead{M5} \\
    \colhead{} & \colhead{} & \colhead{\texttt{pow}} & \colhead{\texttt{cpl}} & \colhead{\texttt{dbb+cpl}} & \colhead{\texttt{dbb+dpbb}} & \colhead{\texttt{dbb+ctt}}
  }
  \startdata
  \texttt{TBabs} & $N_{\rm H}$ ($10^{21}$ cm$^{-2}$)
  & $1.03 \pm 0.12$
  & $1.03_{-0.22}^{+0.12}$
  & $0.90_{-0.34}^{+0.43}$
  & $1.01_{-0.41}^{+0.44}$
  & $0.84_{-0.27}^{+0.33}$ \\
  \hline
  \texttt{diskbb} & $kT_{\rm in,cool}$ (keV)
  & \nodata
  & \nodata
  & $0.27_{-0.06}^{+0.08}$
  & $0.26_{-0.07}^{+0.09}$
  & $0.32_{-0.06}^{+0.07}$ \\
  & Norm
  & \nodata
  & \nodata
  & $2.0_{-1.3}^{+5.4}$
  & $2.0_{-1.4}^{+7.3}$
  & $1.8_{-1.1}^{+1.7}$ \\
  \hline
  \texttt{pow/cpl} & $\Gamma$
  & $1.90 \pm 0.05$
  & $1.90_{-0.20}^{+0.03}$
  & $1.01_{-0.78}^{+0.56}$
  & \nodata
  & \nodata \\
  & $E_{\rm cut}$ (keV)
  & \nodata
  & $>100$
  & $4.4_{-2.0}^{+8.0}$
  & \nodata
  & \nodata \\
  & Norm ($10^{-4}$)
  & $1.41 \pm 0.07$
  & $1.41_{-0.04}^{+0.07}$
  & $1.08_{-0.32}^{+0.23}$
  & \nodata
  & \nodata \\
  \hline
  \texttt{diskpbb} & $T_{\rm in}$ (keV)
  & \nodata
  & \nodata
  & \nodata
  & $2.6_{-0.7}^{+1.2}$
  & \nodata \\
  & $p$
  & \nodata
  & \nodata
  & \nodata
  & $0.58_{-0.04}^{+0.17}$
  & \nodata \\
  & Norm ($10^{-4}$)
  & \nodata
  & \nodata
  & \nodata
  & $2.9_{-2.4}^{+4.5}$
  & \nodata \\
  \hline
  \texttt{compTT} & $kT_{\rm 0}$ (keV)
  & \nodata
  & \nodata
  & \nodata
  & \nodata
  & $0.56_{-0.13}^{+0.16}$ \\
  & $kT_{\rm e}$ (keV)
  & \nodata
  & \nodata
  & \nodata
  & \nodata
  & $2.5$$^{f}$  \\
  & $\tau$
  & \nodata
  & \nodata
  & \nodata
  & \nodata
  & $6.2_{-0.9}^{+0.7}$ \\
  & Norm ($10^{-4}$)
  & \nodata
  & \nodata
  & \nodata
  & \nodata
  & $0.55 \pm 0.08$ \\
  \hline
   & $F_{\rm 0.3-10 keV}^{a}$
  & $0.84 \pm 0.03$
  & $0.84_{-0.07}^{+0.03}$
  & $0.78_{-0.05}^{+0.06}$
  & $0.80_{-0.03}^{+0.04}$
  & $0.77 \pm 0.03$ \\
  & $L_{\rm 0.3-10 keV}^{b}$
  & $1.00 \pm 0.04$
  & $1.00_{-0.08}^{+0.04}$
  & $0.93_{-0.06}^{+0.07}$
  & $0.96_{-0.04}^{+0.05}$
  & $0.92 \pm 0.04$ \\
  & $C$-stat / d.o.f
  & 1939.6/2400
  & 1939.6/2399
  & 1926.7/2397
  & 1927.7/2397
  & 1924.8/2397 \\
  \enddata
  \tablecomments{
    Errors are quoted at 90\% confidence intervals. $^{a}$Unabsorbed flux in the 0.3--10 keV band in units of $10^{-12}$ erg s$^{-1}$ cm$^{-2}$. $^{b}$Unabsorbed luminosity in the 0.3--10 keV band in units of $10^{41}$ erg s$^{-1}$, assuming a distance of 31.6 Mpc. $^{f}$Frozen at the best-fit value of the 2024 epoch. Continuum models: M1: \texttt{TBabs*powerlaw}, M2: \texttt{TBabs*cutoffpl}, M3: \texttt{TBabs*(diskbb+cutoffpl)}, M4: \texttt{TBabs*(diskbb+diskpbb)}, and M5: \texttt{TBabs*(diskbb+compTT)}.
  }
\end{deluxetable*}

For Model M3, the spectral cut-off was detected at a relatively low energy of $E_{\rm cut} = 4.4^{+8.0}_{-2.0}$ keV, consistent with that observed in other ULXs \citep{stobbart2006}. However, given the limited energy bandpass of \textit{XMM-Newton}, the inferred cutoff energy should be interpreted as indicative rather than a robust measurement. In Model M4, the temperature profile exponent was again significantly flatter than that expected for a thin disk ($p\approx 0.75$), with $p = 0.58^{+0.17}_{-0.04}$, and the hot disk temperature was measured to be $T_{\rm in, hot} = 2.6^{+1.2}_{-0.7}$ keV. In both models, the cooler thermal disk or wind photosphere had a temperature of $T_{\rm in,cool} \approx 0.3$ keV.

For Model M5, we initially linked the seed photon temperature to the cool disk temperature, resulting in an acceptable fit (C-stat/d.o.f $\approx 1930/2398$), but the disk normalization was unconstrained. Allowing the seed photon temperature to vary improved the fit in both epochs ($\Delta C \approx 5$ for the 2022 data). Since the plasma temperature was unconstrained, we froze it at $kT_{\rm e} = 2.5$ keV, consistent with the 2024 epoch for direct comparison, assuming the plasma temperature did not change significantly between the epochs. This resulted in a marginal increase in the fit statistic ($\Delta C < 1$) and a well-constrained optical depth of $\tau =6.2_{-0.9}^{+0.7}$. The model preferred a seed photon temperature ($kT_{\rm 0} = 0.56^{+0.16}_{-0.13}$ keV) hotter than the disk temperature ($kT_{\rm in} = 0.32^{+0.07}_{-0.06}$ keV), suggesting a different origin for the seed photons.

\begin{figure}
\epsscale{1.1}
  \plotone{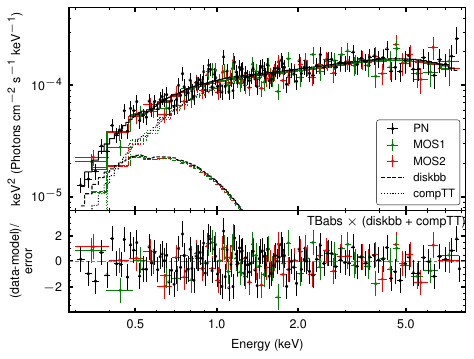}
  \caption{Spectral fit to the 2022 \textit{XMM-Newton} data using Model M5 (\texttt{TBabs*(diskbb+compTT)}). Top panel: Spectrum unfolded through the instrumental response assuming the best-fit model, shown as $E^{2}F(E)$, with the individual model components overplotted: \texttt{diskbb} (dashed) and \texttt{compTT} (dotted). Bottom panel: Residuals with respect to the continuum model. Data in both panels are rebinned for visual clarity.
  \label{fig:spec22}}
\end{figure}

The unabsorbed luminosity in the 0.3--10 keV band for the 2022 epoch was $L_{\rm X} = 9.6^{+0.5}_{-0.4} \times 10^{40}$~erg~s$^{-1}$ (with model M3). Without hard X-ray coverage, the residuals corresponding to the different continuum models were visually indistinguishable, indicating strong spectral degeneracy in the soft X-ray band. However, for direct comparison with the 2024 epoch spectral fits described later, we retain and present the fit results for all considered continuum models in Table~\ref{tab:spectral_fits_2022}. The spectral fit with the best-fit model is shown in Figure~\ref{fig:spec22}.

\subsection{2024: Joint \textit{XMM-Newton} + \textit{NuSTAR} Observation}

For the 2024 epoch, joint spectral fitting was performed in the 0.3--10~keV band for \textit{XMM-Newton} data and the 3.0--15~keV band for \textit{NuSTAR} data, above which the background became dominant. 

Initial fitting with spectra from all instruments (\textit{XMM-Newton}: EPIC-pn, EPIC-MOS1, and EPIC-MOS2; \textit{NuSTAR}: FPMA and FPMB) using the baseline continuum model M3, yielded a statistically acceptable fit ($C$-stat/d.o.f = 3214.8/3590). The resulting best-fit parameters ($N_{\rm H} = (1.0 \pm 0.3) \times 10^{21}$~cm$^{-2}$, $kT_{\rm in} = 0.26_{-0.05}^{+0.06}$~keV, $\Gamma = 1.01_{-0.23}^{+0.19}$, and $E_{\rm cut} = 5.1_{-0.9}^{+1.0}$~keV) were consistent with those obtained from the 2022 epoch within uncertainties (see Table~\ref{tab:spectral_fits_2022}). 
However, upon closer inspection of the fit, we observed pronounced residuals near 2~keV. 

While both the EPIC-pn and EPIC-MOS2 detectors exhibited consistent residuals at this energy, the EPIC-MOS1 detector did not show the feature. The EPIC-pn detector provides the largest effective area and the highest photon statistics ($\approx 8400$ counts). The fact that the feature is independently resolved by EPIC-MOS2 makes the detection robust. We note that EPIC-MOS1 registered $\approx 13\%$ fewer photon counts than EPIC-MOS2 due to a shorter exposure time. It is therefore likely that the MOS1 detector could not resolve the feature due to these lower photon statistics. To avoid artificially diluting the statistical significance of the feature in the joint fits, we excluded the MOS1 spectrum from the primary analysis and performed subsequent spectral modeling using the simultaneous EPIC-pn, EPIC-MOS2, \textit{NuSTAR} FPMA, and \textit{NuSTAR} FPMB spectra.

\begin{figure}
\epsscale{1.1}
  \plotone{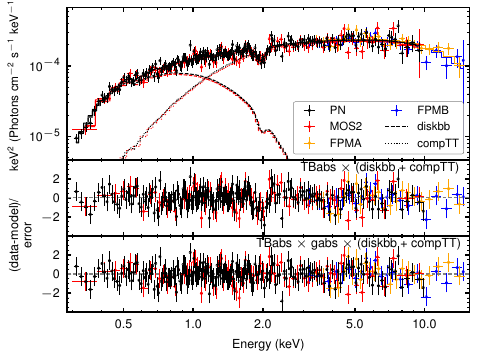}
  \caption{Spectral fit to the 2024 \textit{XMM-Newton}+\textit{NuSTAR} data. Top panel: Spectrum unfolded through the instrumental response assuming the best-fit continuum+line model, \texttt{TBabs*gabs*(diskbb+compTT)}, shown as $E^{2}F(E)$, with the individual continuum components overplotted: \texttt{diskbb} (dashed) and \texttt{compTT} (dotted). Middle panel: Residuals with respect to the continuum-only model, \texttt{TBabs*(diskbb+compTT)}. Bottom panel: Residuals with respect to the continuum+line model, \texttt{TBabs*gabs*(diskbb+compTT)}. Data in all panels are rebinned for visual clarity.}
  \label{fig:spec24}
\end{figure}

For a direct comparison with the 2022 spectral fits, we modeled only the spectral continuum first with the same set of models described in the previous subsection.

The data strongly disfavored an absorbed power-law model (M1) showing significant curvature at higher energies with a $C$-stat/d.o.f of $ 2552.8/2711$. The cutoff power-law model (M2) yielded a well-constrained cut-off energy $E_{\rm cut}=8.2_{-1.3}^{+1.9}$ keV and a photon index of $\Gamma_{\rm cpl}=1.40 \pm 0.08$ (\textit{C}-stat/d.o.f $= 2472.9/2710$). 

\begin{deluxetable*}{llccccc c ccc}
  \tabletypesize{\scriptsize}
  \setlength{\tabcolsep}{2pt}
  \tablecaption{Best-fit spectral parameters for the simultaneous 2024 \textit{XMM-Newton} + \textit{NuSTAR} observation (ObsID 0931440101 + 50901001002). 
  \label{tab:spectral_fits_2024}}
  \tablehead{
    \colhead{Component} & \colhead{Parameter} & \multicolumn{5}{c}{Continuum-only Model} & \colhead{} & \multicolumn{3}{c}{Continuum + Line Model} \\
    \cline{3-7} \cline{9-11}
    \colhead{} & \colhead{} & \colhead{M1} & \colhead{M2} & \colhead{M3} & \colhead{M4} & \colhead{M5} & \colhead{} & \colhead{\texttt{gabs}*M3} & \colhead{\texttt{gabs}*M4} & \colhead{\texttt{gabs}*M5} \\
    \colhead{} & \colhead{} & \colhead{\texttt{pow}} & \colhead{\texttt{cpl}} & \colhead{\texttt{dbb+cpl}} & \colhead{\texttt{dbb+dpbb}} & \colhead{\texttt{dbb+ctt}} & \colhead{} & \colhead{} & \colhead{} & \colhead{}
  }
  \startdata
  \texttt{TBabs} & $N_{\rm H}$ ($10^{21}$ cm$^{-2}$) & $1.30 \pm 0.12$ & $0.75_{-0.14}^{+0.15}$ & $0.86_{-0.25}^{+0.31}$ & $1.03_{-0.27}^{+0.33}$ & $0.71_{-0.21}^{+0.25}$ & & $0.94_{-0.34}^{+0.48}$ & $1.17_{-0.34}^{+0.56}$ & $0.88_{-0.24}^{+0.48}$ \\
  \hline
  \texttt{gabs} & $E_{\rm line}$ (keV) & \nodata & \nodata & \nodata & \nodata & \nodata & & $1.966_{-0.039}^{+0.044}$ & $1.964_{-0.039}^{+0.042}$ & $1.966_{-0.039}^{+0.042}$ \\
  & $\sigma_{\rm line}$ (keV) & \nodata & \nodata & \nodata & \nodata & \nodata & & $0.074_{-0.044}^{+0.036}$ & $0.076_{-0.041}^{+0.036}$ & $0.076_{-0.043}^{+0.036}$ \\
  & Strength (keV) & \nodata & \nodata & \nodata & \nodata & \nodata & & $0.078_{-0.027}^{+0.032}$ & $0.080_{-0.027}^{+0.030}$ & $0.083_{-0.030}^{+0.035}$ \\
  & EW$_{\rm line}$ (keV) & \nodata & \nodata & \nodata & \nodata & \nodata & & $-0.067^{+0.027}_{-0.028}$ & $-0.070^{+0.027}_{-0.041}$ & $-0.071^{+0.023}_{-0.042}$ \\
  \hline
  \texttt{diskbb} & $kT_{\rm in,cool}$ (keV) & \nodata & \nodata & $0.29_{-0.06}^{+0.07}$ & $0.27_{-0.08}^{+0.09}$ & $0.41\pm 0.08$ & & $0.24_{-0.09}^{+0.07}$ & $0.19 \pm 0.13$ & $0.29\pm 0.11$ \\
  & Norm & \nodata & \nodata & $1.1_{-0.6}^{+2.1}$ & $1.1_{-0.8}^{+4.4}$ & $0.66_{-0.34}^{+0.74}$ & & $1.5_{-1.3}^{+13.8}$ & $3.0 (< 90)$ & $2.2_{-1.9}^{+3.5}$ \\
  \hline
  \texttt{pow/cpl} & $\Gamma_{cpl}$ & $1.82_{-0.04}^{+0.04}$ & $1.40 \pm 0.08$ & $0.92_{-0.29}^{+0.23}$ & \nodata & \nodata & & $1.17_{-0.29}^{+0.20}$ & \nodata & \nodata \\
  & $E_{\rm cut}$ (keV) & \nodata & $8.2_{-1.3}^{+1.9}$ & $4.84_{-1.0}^{+1.3}$ & \nodata & \nodata & & $5.88_{-1.3}^{+1.7}$ & \nodata & \nodata \\
  & Norm ($10^{-4}$) & $1.60 \pm 0.06$ & $1.57 \pm 0.06$ & $1.18_{-0.23}^{+0.20}$ & \nodata & \nodata & & $1.45_{-0.27}^{+0.19}$ & \nodata & \nodata \\
  \hline
  \texttt{diskpbb} & $T_{\rm in,hot}$ (keV) & \nodata & \nodata & \nodata & $3.01_{-0.30}^{+0.35}$ & \nodata & & \nodata & $3.18_{-0.29}^{+0.35}$ & \nodata \\
  & $p$ & \nodata & \nodata & \nodata & $0.58_{-0.02}^{+0.04}$ & \nodata & & \nodata & $0.56_{-0.01}^{+0.03}$ & \nodata \\
  & Norm ($10^{-4}$) & \nodata & \nodata & \nodata & $2.2_{-1.0}^{+1.9}$ & \nodata & & \nodata & $1.5_{-0.6}^{+1.1}$ & \nodata \\
  \hline
  \texttt{compTT} & $kT_0$ (keV) & \nodata & \nodata & \nodata & \nodata & $0.74_{-0.21}^{+0.20}$ & & \nodata & \nodata & $0.45_{-0.15}^{+0.39}$ \\
  & $kT_{e}$ (keV) & \nodata & \nodata & \nodata & \nodata & $3.1_{-0.8}^{+24.4}$ & & \nodata & \nodata & $2.53_{-0.28}^{+0.62}$ \\
  & $\tau$ & \nodata & \nodata & \nodata & \nodata & $5.2_{-4.9}^{+2.0}$ & & \nodata & \nodata & $6.7_{-1.9}^{+0.9}$ \\
  & Norm ($10^{-5}$) & \nodata & \nodata & \nodata & \nodata & $4.8_{-4.5}^{+2.5}$ & & \nodata & \nodata & $8.2_{-4.5}^{+3.8}$ \\
  \hline
   & $F_{\rm 0.3-10 keV}^a$ & $1.01 \pm 0.02$ & $0.92 \pm 0.03$ & $0.94^{+0.05}_{-0.04}$ & $0.97^{+0.06}_{-0.04}$ & $0.92^{+0.04}_{-0.03}$ & & $0.96^{+0.09}_{-0.05}$ & $1.01 \pm 0.03$ & $0.95^{+0.06}_{-0.05}$ \\
  & $L_{\rm 0.3-10 keV}^b$ & $1.21 \pm 0.03$ & $1.10 \pm 0.03$ & $1.13^{+0.06}_{-0.05}$ & $1.21^{+0.03}_{-0.04}$ & $1.15^{+0.05}_{-0.10}$ & & $1.13^{+0.06}_{-0.05}$ & $1.15 \pm 0.03$ & $1.06 \pm 0.02$ \\
  & $C$-stat / d.o.f & 2552.8/2711 & 2472.9/2710 & 2449.1/2708 & 2450.6/2708 & 2449.1/2707 & & 2424.8/2705 & 2424.6/2705 & 2426.7/2704 \\
  \enddata
  \tablecomments{Errors are quoted at 90\% confidence intervals. $^{a}$Unabsorbed flux in the 0.3--10 keV band in units of $10^{-12}$ erg s$^{-1}$ cm$^{-2}$. $^{b}$Unabsorbed luminosity in the 0.3--10 keV band in units of $10^{41}$ erg s$^{-1}$, assuming a distance of 31.6~Mpc. Continuum models: M1: \texttt{TBabs*powerlaw}, M2: \texttt{TBabs*cutoffpl}, M3: \texttt{TBabs*(diskbb+cutoffpl)}, M4: \texttt{TBabs*(diskbb+diskpbb)}, and M5: \texttt{TBabs*(diskbb+compTT)}. In \texttt{gabs}*M4, the \texttt{diskbb} norm was unconstrained and the best-fit value is provided with 90\% confidence upper limits in parentheses. MOS1 data is excluded from spectral analysis. See text for more details.}
\end{deluxetable*}

The addition of the thermal \texttt{diskbb} component to the models M3, M4, and M5 yielded statistically significant improvements, with $\Delta C$ of $-23.8$, $-11.6$, and $-8.1$, respectively, for 2 additional d.o.f.

Model M3 indicated a spectral energy cut-off at $E_{\rm cut}=4.8_{-1.0}^{+1.3}$~keV with a hard photon index of $\Gamma_{\rm cpl} =0.92_{-0.29}^{+0.23}$ along with a thermal component temperature of $kT_{\rm in,cool}=0.29^{+0.07}_{-0.06}$~keV, closely matching the spectral parameters of the known pulsating ULX population \citep{pintore2017}. 

Model M4 estimated a thermal component temperature, $kT_{\rm in,cool} = 0.27^{+0.09}_{-0.08}$ keV while providing well-constrained values for the inner hot disk temperature, $kT_{\rm in, hot} = 3.01_{-0.30}^{+0.35}$ keV, and the radial temperature profile exponent, $p=0.58_{-0.02}^{+0.04}$ consistent with an advection-dominated disk supporting the idea of a super-Eddington accretor in the system. We note that adding a third component, such as \texttt{cutoffpl} in continuum model M4 with fixed $\Gamma_{\rm cpl}=0.59$ and $E_{\rm cut}=7.9$~keV (the magnetic accretor model; see \citealt{walton2018,walton2020,walton2025}), resulted in only a small improvement in the $C$-statistic ($\Delta C \approx 3.1$).

Model M5 poorly constrained the plasma temperature and optical depth at $kT_{e} = 3.1^{+24.4}_{-0.8} $~keV and $ \tau = 5.2^{+2.0}_{-4.9} $, respectively, prior to modeling the residuals near 2 keV. We obtained a seed photon temperature of $kT_{\rm 0} = 0.74^{+0.20}_{-0.21}$~keV and a thermal component temperature of $kT_{\rm in,cool} = 0.41 \pm 0.08$~keV.

We present the best-fit spectral parameters for all considered continuum models in Table~\ref{tab:spectral_fits_2024}. Notably, the unabsorbed luminosity in the 0.3--10 keV energy band for the 2024 epoch was $L_{\rm X} = 1.13^{+0.06}_{-0.05} \times 10^{41}$~erg~s$^{-1}$ (with model M3), confirming its hyperluminous state.

\subsection{Detection of a 1.97 keV Absorption Feature}
\label{subsec:absfeat}

The residual near 2~keV was well modeled using a Gaussian profile with the \texttt{gabs} model component in \texttt{XSPEC}. The addition of this Gaussian absorption line to the dual-component models significantly improved the fit, with an observed reduction in $C$-stat ($\Delta C$) of $-24.34$, $-25.99$, and $-22.40$ for 3 additional degrees of freedom for models M3, M4, and M5, respectively. For model M3, the improvement is driven primarily by the PN and MOS2 spectra, which contribute $\Delta C_{\rm PN} = -17.64$ and $\Delta C_{\rm MOS2} = -7.37$, while the contribution from the \textit{NuSTAR} detectors is negligible since they do not cover the energy range of the feature. The absorption line parameters were well-constrained, with the line energy centered at $E_{\rm line} = 1.966^{+0.044}_{-0.039}$~keV, a line width of $\sigma_{\rm line} = 74^{+36}_{-44}$~eV, a line strength of $78^{+32}_{-27}$~eV, and an equivalent width of $EW_{\rm line} = -67^{+27}_{-28}$~eV (for model M3). The feature was robust against the choice of continuum model, and the line parameters were consistent across all models. We note that the best-fit line centroid energy is well-separated from known instrumental artifacts in the soft X-ray band, sitting cleanly between the \textit{XMM-Newton} detector Si K-edge ($\sim$1.84 keV) and the Au M-edge of the \textit{XMM-Newton} mirrors ($\sim$2.3 keV). We also ruled out contamination from the nearby LLAGN ($\sim 1\arcmin$ away; see Appendix \ref{app:agn}). 

In order to search for the absorption feature in the 2022 epoch, we specifically added a \texttt{gabs} component to the 2022 \textit{XMM-Newton} spectral continuum, with the line centroid energy fixed at 1.97~keV and the line width fixed at 75~eV. However, the line strength could not be constrained, and we therefore treated this as a non-detection of the feature. The 90\% confidence upper limit on the \texttt{gabs} strength parameter was 0.52~keV.

\begin{figure}
\epsscale{1.1}
\gridline{\fig{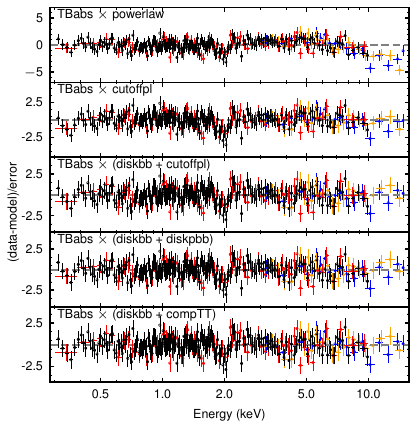}{0.47\textwidth}{(a)}}
\vspace{-10pt}
\gridline{
\fig{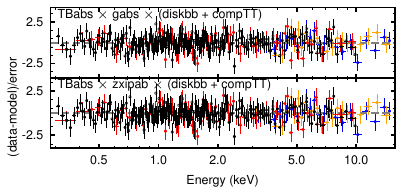}{0.47\textwidth}{(b)}
}
\caption{Spectral residuals from fitting the 2024 simultaneous \textit{XMM-Newton}+\textit{NuSTAR} data in the 0.3--15 keV band. Data from EPIC-pn (black), EPIC-MOS2 (red), FPMA (orange), and FPMB (blue) are plotted. Panel (a) shows the residuals for the models M1-M5. Panel (b) shows the residuals for the models \texttt{TBabs*gabs*(diskbb+compTT)} and \texttt{TBabs*zxipab*(diskbb+compTT)}, which account for the absorption feature.
\label{fig:resid24_7}}
\end{figure}

The inclusion of the Gaussian absorption line to the 2024 epoch joint-spectra did not significantly affect the other parameter values, which remained consistent with the continuum-only spectral fit results within their respective uncertainties. In Model M3, the spectral cutoff energy and photon index were found to be $E_{\rm cut}=5.9_{-1.3}^{+1.7}$~keV and $\Gamma_{\rm cpl}=1.17_{-0.3}^{+0.2}$, respectively. While the \texttt{diskbb} norm was unconstrained at the 90\% confidence level for Model M4, the inner hot disk parameters remained consistent with the previous fits. For Model M5, the plasma parameters became well-constrained ($\tau=6.7^{+0.9}_{-1.9}$ and $kT_{e}=2.53_{-0.28}^{+0.62}$~keV), with a seed photon temperature of $kT_{0}= 0.45^{+0.39}_{-0.15}$~keV and a soft thermal component temperature of $kT_{\rm in, cool} = 0.29\pm 0.11$~keV. The best-fit parameters are presented in the \textit{Continuum+Line} section of Table \ref{tab:spectral_fits_2024}. We plot the spectral fit to the data using Model M5 in Figure \ref{fig:spec24}, and show the residuals from the fits for all considered models in Figure~\ref{fig:resid24_7}.

\begin{figure}
\epsscale{1.2}
\plotone{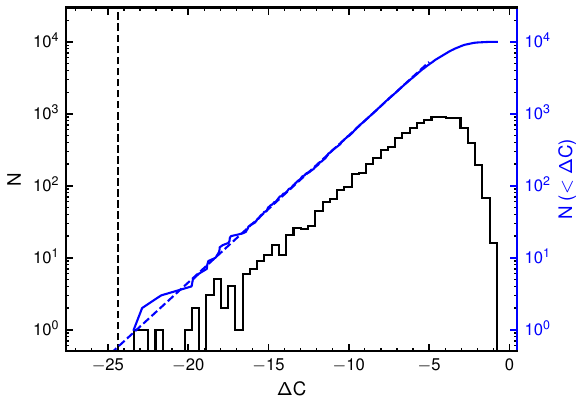}
\caption{Distribution of the maximum improvement in the fit statistic ($\Delta C$) obtained from $10,000$ simulated spectra generated from the continuum-only model M3. The vertical, dashed, black line indicates the observed improvement ($\Delta C = -24.34$) from the real data. None of the simulated random fluctuations in the continuum produced a $\Delta C$ greater than the observed value, implying a False Alarm Rate (FAR) of $< 10^{-4}$ and a detection significance of $>3.9\sigma$. The Cumulative Distribution Function (CDF) of $\Delta C$ (solid blue curve) is overlaid, showing the fitted line (dashed, blue line) used to extrapolate the tail probability.}
\label{fig:Cdist}
\end{figure}

We performed extensive Monte Carlo simulations to determine the detection significance of the line, following the method described in \cite{bright2022}, to account for the look-elsewhere effect (LEE; see \citealt{pinto2023}). $10,000$ synthetic source spectra were generated using the \texttt{fakeit} command in \texttt{XSPEC} for each of the continuum-only models M3, M4, and M5 without the line component, using the same observed background and instrument responses. The simulated spectra were then refitted with respective continuum-only models before adding \texttt{gabs}. We used the \texttt{steppar} command to perform a Gaussian line scan between $1.0$--$8.0$ keV in 100 linearly spaced steps. At each energy step, the simulated spectra were fitted with the Gaussian line component width fixed at $75$ eV, and the line strength left to vary. The largest improvement in the $C$-statistic ($\Delta C$) obtained from adding the line component at any energy step in the defined range was recorded for each simulation. We plot the resulting distribution of these maximum $\Delta C$ values (for model M3) in Figure~\ref{fig:Cdist}. 

For all three models, after $10,000$ simulations, none of the $\Delta C$ values derived from random fluctuations exceeded the observed improvement in $C$-statistic (e.g., $\Delta C = -24.34$ for M3). This suggests a False Alarm Rate (FAR) of $<10^{-4}$, which corresponds to a two-sided detection significance of $>3.9 \sigma$.

A larger number of simulations ($>10^{5}$) is required to measure a high detection significance ($>3.9 \sigma$). However, we employed an exponential tail extrapolation method (see \citealt{bright2022}) to obtain a reliable estimate. A cumulative distribution function (CDF) of maximum $\Delta C$ values, $N(<\Delta C)$, was generated (see Figure~\ref{fig:Cdist}). The tail of this distribution follows an exponential decay, but in the semi-logarithmic space takes a linear form. A function given by $\log_{10}(N) = m \cdot \Delta C + b$, was used to fit the linear section of the CDF between $\Delta C = -7$ and $\Delta C = -15$, where $m$ is the slope and $b$ is the intercept. The resulting slope and intercept were $m=0.204$ and $b=4.74$, respectively (for model M3). Given the observed $\Delta C=-24.34$, the false alarm count was calculated to be 0.587, corresponding to an FAR of $5.87\times 10^{-5}$ when normalized by the total number of simulations. This indicates a detection significance for the line of $\approx 4.0 \sigma$. The corresponding extrapolated detection significances for models M4 and M5 are $\approx 4.2 \sigma$ and $\approx 3.9 \sigma$, respectively.

\begin{figure}
\epsscale{1.2}
\plotone{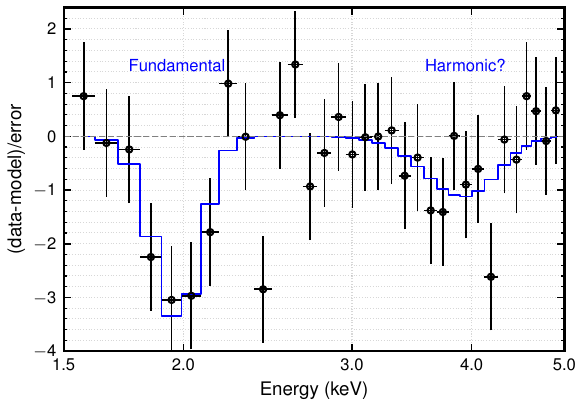}
\caption{Residuals in EPIC-pn spectrum after fitting continuum model \texttt{TBabs*(diskbb+compTT)}. The Gaussian absorption line at $\approx 1.97$ keV is marked as the fundamental, with the potential harmonic centered at $\sim 4$ keV. The blue curves represent the best-fit Gaussians to the residual features. The data has been visually rebinned to have at least 15$\sigma$ significance and a minimum of 20 counts per bin.
\label{fig:residline}}
\end{figure}

For completeness, we repeated the analyses including the EPIC-MOS1 spectrum. When all five spectra (EPIC-pn, EPIC-MOS2, EPIC-MOS1, \textit{NuSTAR} FPMA, and \textit{NuSTAR} FPMB) were used, the addition of the Gaussian absorption line component improved the fit statistic from $C\text{-stat/d.o.f} = 3214.76/3590$ to $C\text{-stat/d.o.f} = 3198.22/3587$ (for model M3), corresponding to a $\Delta C = -16.54$ for three additional d.o.f. EPIC-pn and EPIC-MOS2 showed an improvement of $\Delta C = -14.57$ and $\Delta C = -6.99$, respectively, while EPIC-MOS1 exhibited a modest worsening of $\Delta C = +4.42$, indicating that it does not independently require the line component. The line centroid energy was poorly constrained with $E_{\rm line}=1.93\pm0.48$~keV, a line width of $\sigma_{\rm line}=0.080 \pm 0.045$~keV, and a line strength of $S=0.057^{+0.027}_{-0.025}$ keV. Repeating the Monte Carlo line-scan procedure with 10,000 simulated spectra produced $17$ false alarms, corresponding to an FAR of $1.7 \times 10^{-3}$ and a detection significance of $\approx 3.1 \sigma$.

\subsection{Searching for Cyclotron Harmonics}

We observed a broad and shallow residual feature centered around 4 keV, most pronounced in the EPIC-pn spectrum in the 2024 epoch (see Figure \ref{fig:residline}) for all the continuum fits. Since cyclotron features are expected to be accompanied by harmonics at approximately integer multiples of the fundamental energy, we attempted to model this residual with a second \texttt{gabs} component. To reduce parameter degeneracy, we fixed the value of the line energy to twice that of the fundamental energy ($E_{\rm line,2}=3.94$ keV). The addition of the second line resulted in a marginal improvement of $\Delta C = -4.64$ for 2 additional degrees of freedom relative to the single-line model. The best-fit width of the second feature was broad, $\sigma_{\rm line,2} = 0.30_{-0.19}^{+0.61}$~keV, with a line strength of $0.09_{-0.07}^{+0.14}$~keV. We note that this behaviour was consistent across continuum models.

We removed the two Gaussian line components and replaced them with a single physically motivated \texttt{cyclabs} model \citep{mihara1990, makishima1990}, which uses two pseudo-Lorentzian profiles to simultaneously fit both the fundamental cyclotron feature and its harmonic. The depth parameters ($D_f, D_{h}$) in this model quantify the magnitude of continuum attenuation at each resonance energy. With the fundamental energy fixed at 1.97~keV, the fit resulted in a $C$-stat = 2423.4 (for 2704~d.o.f.), corresponding to only a marginal improvement of $\Delta C = -1.43$ for one additional degree of freedom relative to the single \texttt{gabs} model, indicating that the residual near 4~keV was not statistically required. While the harmonic depth was constrained to $D_{h} = 0.14_{-0.06}^{+0.61}$, both the fundamental and harmonic widths remained unconstrained. We concluded from the above analysis that the residuals near 4 keV were suggestive but statistically insignificant and were consistent with a non-detection of the harmonic.

\begin{figure}
\epsscale{1.2}
\plotone{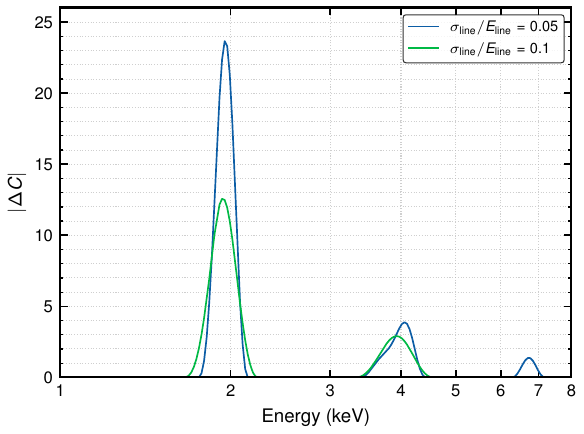}
\caption{The improvement in the C-statistic ($|\Delta C|$) resulting from a blind Gaussian line scan across the joint spectral continuum as a function of energy. The scan used Gaussian absorption profiles with fractional widths ($\sigma_{\rm line}/E_{\rm line}$) fixed at $0.05$ and $0.1$ to represent line morphologies typically associated with proton and electron CRSFs, respectively.
\label{fig:cstatvsE}}
\end{figure}

In Figure \ref{fig:cstatvsE}, we show the improvement in the fit statistic resulting from a blind Gaussian line scan across the joint spectral continuum (using model M3). The scan was performed by adding a \texttt{gabs} component in \texttt{XSPEC} and stepping the line centroid energy across a logarithmic array between 1 and 8 keV. To account for the different line morphologies typically associated with proton and electron CRSFs, the line scan was carried out for two fixed fractional widths: a narrow profile ($\sigma_{\rm line}/E_{\rm line} = 0.05$) and a broader profile ($\sigma_{\rm line}/E_{\rm line} = 0.1$). The scan recovered the highly significant detection of the absorption feature at $E \approx 1.97$ keV, with a maximum improvement of $\Delta C \approx 23.7$. Furthermore, a marginal improvement was observed near $\sim$4~keV ($\Delta C \approx 4$), which lies exactly at the first harmonic interval assuming the 1.97 keV feature represents a fundamental CRSF. Notably, the scan also revealed a small improvement near 6.7~keV, although this feature was not statistically significant, and we therefore do not attempt a physical interpretation.

\subsection{Testing Photoionized Absorption Models}
\label{sec:photmod}

ULXs are known to launch powerful, large-scale radiation-driven winds which are predicted to reach relativistic speeds of 0.1--0.3$c$ \citep{ohsuga2007,pinto2017,kosec2018}. To test the possibility of the absorption feature originating from an ionized, fast outflow, we first modeled it with the partial-covering ionized absorber model, \texttt{zxipcf} \citep{reeves2008}, which has been used to explain absorption features in soft X-ray spectra of super-Eddington accretors \citep[e.g.,][]{luangtip2015,lin2017}.

The continuum model M5 modified by this absorber, \texttt{TBabs*zxipcf*(diskbb+compTT)}, yielded a statistically acceptable fit with $C\text{-stat/d.o.f.} = 2435.4/2703$, requiring an extreme column density of $N_{\rm H} = 1.80_{-0.95}^{+0.66} \times 10^{24}\ \mathrm{cm^{-2}}$, a covering fraction of $f = 0.66_{-0.29}^{+0.31}$, and an ionization parameter of $\log \xi = 3.23_{-0.27}^{+0.61}$. The model attempted to fit the discrete 1.97 keV feature by applying a blueshift of $z = -0.437_{-0.024}^{+0.008}$ to the atomic transitions, which corresponds to a line-of-sight outflow velocity of $v_{\rm LOS} = 0.52_{-0.01}^{+0.03}c$ using the relativistic Doppler shift relation. Such an extreme outflow velocity is difficult to reconcile with current models of radiatively driven ULX winds \citep{ohsuga2007,king2016}.

Since super-Eddington winds in ULXs are expected to be highly clumpy, stratified, and multi-phase \citep{takeuchi2013, kobayashi2018,pinto2021}, a single-zone model like \texttt{zxipcf}, which assumes a single column density, would likely be insufficient to model the wind structure. Therefore, we used the \texttt{zxipab} model component \citep{nazma2021} in \texttt{XSPEC} as a proxy, which uses the same \texttt{XSTAR} photo-ionisation grids as \texttt{zxipcf} while incorporating the \texttt{pwab} model to apply a continuous, power-law distribution of covering fractions as a function of column density.

To prevent the \texttt{TBabs*zxipab*(diskbb+compTT)} model from taking unphysical values, we froze the minimum equivalent hydrogen column density, $n_{\rm H,min}$, to a fiducial $1 \times 10^{21}\ \mathrm{cm^{-2}}$, corresponding to the established best-fit neutral hydrogen column density for this source, and the power-law index to $\beta=0$, forcing the model to integrate over a uniform distribution of gas. This resulted in a statistically acceptable fit ($C$-stat/d.o.f. $= 2428.1/2704$) comparable to the single Gaussian line model. We show the residuals from the spectral fit with this model in the bottom panel of Figure \ref{fig:resid24_7}b. The free parameters of the \texttt{zxipab} model were constrained with a maximum equivalent hydrogen column density, $n_{\rm H,max}=1.25^{+5.12}_{-0.54} \times 10^{22}\ \mathrm{cm^{-2}}$ and an ionization parameter of $\log \xi = 2.73^{+0.28}_{-0.39}$. The model required a blueshift of the ionized absorber of $z=-0.310 \pm 0.005$, which corresponds to an outflow velocity of $v_{\rm LOS} \approx 0.36c$. This velocity represents a lower limit on the intrinsic outflow speed, since the observed blueshift measures only the line-of-sight component. Any inclination of the wind relative to the observer's line-of-sight would imply an even higher intrinsic velocity. We note that the inferred absorber blueshift remained insensitive to the frozen parameters.

\subsection{Spectral Evolution between 2022 and 2024}
\label{sec:specevol}

\begin{figure}
\centering
\gridline{\fig{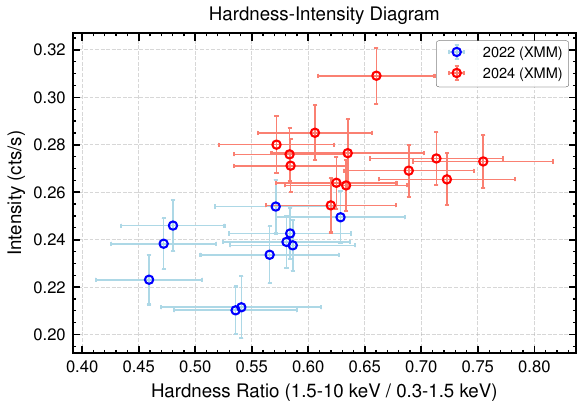}{0.48\textwidth}{(a)}} 
\gridline{\fig{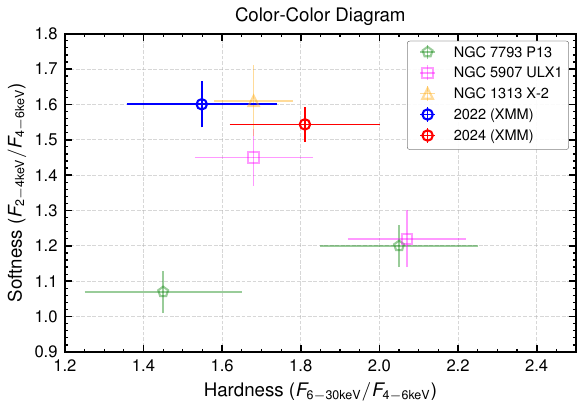}{0.48\textwidth}{(b)}} 
\caption{(a) Hardness--Intensity Diagram (HID) of NGC~3583~X-1 comparing the 2022 (blue) and 2024 (red) epochs. (b) Color Color Diagram (CCD) of the source comparing its hardness and softness during the 2022 (blue) and 2024 (red) epochs with known PULXs NGC~5907~ULX1 (magenta), NGC~7793~P13 (green) and NGC~1313~X-2 (orange). All error bars correspond to 1$\sigma$ uncertainties. See text for details. 
\label{fig:hidccd}} 
\end{figure}

We plotted a Hardness--Intensity diagram (HID) (Figure \ref{fig:hidccd}a) from the \textit{XMM-Newton} EPIC-pn data using 3~ks time bins. We defined hardness as ratio of count rates in the hard band (1.5--10~keV) to that in the soft (0.3--1.5~keV) band and the intensity as the total count rate in the full 0.3--10 keV band. The source showed a clear brighter-when-harder trend from the 2022 to 2024 epochs. 

To allow for a direct comparison with other ULX studies \citep[e.g.,][]{pintore2017}, we used the phenomenological model M3 for the 2022 epoch and the \texttt{gabs}*M3 model for the 2024 epoch to account for the detected absorption feature. We plotted a Color-Color Diagram (CCD) of the source using only the \textit{XMM-Newton} data, comparing its hardness (ratio of model-predicted fluxes in the 6--30 keV and 4--6 keV band) and softness (ratio of model-predicted fluxes in the 2--4 keV and 4--6 keV band) during the 2022 and 2024 epochs, with known PULXs NGC~5907~ULX1, NGC~7793~P13 and NGC~1313~X-2 (see Figure \ref{fig:hidccd}b).

In the 2022 epoch, we obtained a softness of $1.60 \pm 0.07$ and a hardness of $1.55 \pm 0.19$. Although the source became brighter in the 2024 epoch, the softness ($1.54 \pm 0.05$) and the hardness ($1.81 \pm 0.19$) remained consistent within uncertainties (quoted at the $1\sigma$ level). Notably, the source occupied a region of the CCD that is typically associated with the known PULX sample (see Figure 2 of \citealt{pintore2017}).

To determine the spectral state of the source during the two epochs according to the empirical ULX scheme defined by \citet{sutton2013}, we used an absorbed disk plus power-law model to fit only the \textit{XMM-Newton} data. In both epochs, the disk temperature was found to be $<0.5$~keV ($kT_{\rm in} \approx 0.21$ keV and $0.15$ keV in 2022 and 2024, respectively) and the power-law photon index remained hard with $\Gamma < 2.0$ ($\Gamma \approx 1.84$ and $1.75$ in 2022 and 2024, respectively). Thus, the source was firmly placed in the Hard Ultraluminous (HUL) state during both epochs.

\section{Discussion}
\label{sec:disc}

\subsection{Broadband Spectral modeling of NGC 3583 X1}
\label{sec:contmod}

Using the phenomenological model M3, we obtained a clear spectral energy cut-off at $\sim 5$--$6$~keV, a feature expected for the super-Eddington accretion regime \citep{stobbart2006}. Furthermore, Model~M4 produced a radial temperature profile exponent of $p \approx 0.56$, representing a significant deviation from the $p=0.75$ expected for a standard thin disk. Such flattened profiles are widely interpreted as signatures of a puffed-up, advection-dominated inner accretion disk operating in the super-Eddington regime \citep{abramowicz1988, ghosh2021, ghosh2023}, which strongly disfavors a sub-Eddington IMBH interpretation. Additionally, the relatively cool ($kT_e \approx 2.5$~keV) and optically thick ($\tau \approx 6$--$7$) Comptonizing plasma found in Model~M5 is characteristic of the ``ultraluminous state'' observed in super-Eddington ULXs \citep{gladstone2009}.

NGC~3583~X-1 exhibited a harder-when-brighter behavior in the HID and occupied a position on the CCD typically populated by PULXs. Moreover, it was firmly placed in the HUL state during both epochs, which typically implies a low-inclination or near face-on viewing geometry \citep{sutton2013}.

We can estimate the inner disk emitting radius $R_{\rm in}$ using the normalization of the hard component ($N_{\rm dpbb}=1.5_{-0.6}^{+1.0}\times10^{-4}$) from the \texttt{gabs}*M4 model. Assuming a face-on geometry (implied by the HUL state), a standard color correction factor ($\kappa \approx 3$, \citealt{watarai2003}) and an inner boundary condition correction factor ($\xi=0.353$, \citealt{vierdayanti2008}), we obtained an inner disk radius of $R_{\rm in} \approx 123^{+36}_{-28}$~km using the relation $R_{\rm in} = \xi \kappa^{2} \sqrt{N_{\rm dpbb}} D_{10\text{ kpc}}$ \citep{kubota1998}, where $D_{10\text{ kpc}}$ is the distance to the source in units of 10~kpc. If we interpret this radius as the innermost stable circular orbit (ISCO) of a non-rotating BH ($R_{\rm ISCO} = 6GM_{\rm BH}/c^{2}$), it would imply a mass of $M_{\rm BH} \approx 13.9^{+4.0}_{-3.1}\ M_{\odot}$, effectively ruling out an IMBH interpretation.

\subsection{Is NGC~3583~X-1 a Neutron Star Accretor?}

We primarily explored two physical interpretations for the 1.97 keV absorption feature typically invoked for the origin of such features in the ULX spectra: ultrafast outflows (UFOs) and cyclotron resonance scattering features (CRSFs). We discuss the two scenarios in the following subsections.

\subsubsection{Arguments against a UFO interpretation}

High-resolution spectroscopy of ULXs using \textit{XMM-Newton}/RGS data by \citet{kosec2021} has shown that absorption features in ULXs are predominantly detected in sources exhibiting the soft ultraluminous (SUL) state and are either weak or absent in the hard ultraluminous (HUL) state. Although winds cannot be excluded in the HUL state, the nearly face-on viewing geometry generally associated with HUL sources implies that the line of sight is expected to intersect a lower column of outflowing material, making strong absorption features less likely. We note that, while \textit{XMM-Newton}/RGS data were available for both epochs, the RGS spectra were overwhelmingly background dominated, and no meaningful constraints could be placed on any of the atomic features.

If the absorption feature originates in an ionized outflow, spectral modeling with an ionized absorber (Section \ref{sec:photmod}) suggests a line-of-sight outflow velocity of $v_{\rm LOS} \approx 0.36c$. 
Such an extreme outflow velocity is difficult to reconcile with current radiatively driven super-Eddington wind models, which generally predict terminal velocities of $\sim$0.2--0.3$c$ \citep{ohsuga2007,king2016}. Since ULX winds are expected to be launched away from the polar axis \citep{ohsuga2011,middle2015a,kobayashi2018}, projection effects in the nearly face-on HUL state imply an even higher intrinsic velocity. Together, these considerations suggest that a significant fraction of the available accretion power would need to be channeled into the outflow in mechanical form, making such a wind interpretation difficult to reconcile with standard super-Eddington scenarios.

Notably, an 8.56 keV absorption feature was reported by \citet{bright2022} for the HLX in NGC~4045. Although the authors favored a UFO interpretation, they did not rule out a CRSF interpretation since their model required an extremely high ionization parameter ($\log \xi \approx 4.7$) and column density ($>10^{24}$ cm$^{-2}$). In contrast, our model requires a moderate ionization ($\log \xi \approx 2.75$) and an optically thin column density ($N_{\rm H} \sim 1.3 \times 10^{22}$ cm$^{-2}$). Given the inferred blueshift ($z=-0.31$), the observed 1.97 keV absorption feature would correspond to a rest-frame energy of $\sim 1.33$--$1.35$ keV, consistent with the Mg~\textsc{xi} triplet. Although our photoionized absorber model achieved a statistically comparable fit to the single Gaussian absorption line model, the best-fit solution converged toward a relatively low column density, suppressing accompanying atomic transitions (such as those from Mg~\textsc{xii} and Si) expected from a physically self-consistent ultra-fast outflow. In such a scenario, additional absorption features at corresponding blueshifted energies with comparable strengths would generally be expected \citep{cruz2026}, yet no statistically significant companion transitions were detected in the spectra.

Due to the absence of expected accompanying atomic absorption features, the extreme outflow velocity requirement, and the inferred viewing geometry we find a UFO origin for the line less likely.

\subsubsection{Interpretation as a CRSF and constraints on the Magnetic Field Strength}
\label{subsec:magrad}

A more natural explanation for the isolated 1.97 keV absorption feature is the resonant scattering of photons by charged particles in the presence of a strong magnetic field. This scenario is physically consistent with the HUL state viewing geometry which provides an unobstructed line-of-sight to the inner accretion column where a CRSF would form. If the feature is indeed a CRSF, it would provide strong evidence for the underlying compact object to be a highly magnetized neutron star.

Although statistically insignificant, the residuals near 4 keV ($\approx 2 \times E_{fund}$) could be interpreted as a suppressed harmonic which would further support a CRSF interpretation. The strength of cyclotron harmonics strongly depends on the viewing geometry and the relative angle between the line-of-sight and the magnetic field \citep{daugherty1977,harding1991,meszaros1992}. The harmonic can be significantly suppressed or appear shallow in phase-averaged spectra due to the superposition of a large number of lines formed at different heights within the accretion column where the local magnetic field strength varies \citep{nishimura2015}. In galactic accreting pulsars, harmonic CRSFs have been observed to be undetectable in the phase-averaged spectra due to this smearing effect \citep{enoto2008,molkov2019}. Therefore, the absence of a significant harmonic in our phase-averaged spectrum does not rule out its existence.

An absorption feature was not observed in the 2022 epoch spectrum when the source was fainter. However, the 1.97 keV feature was significantly detected in the 2024 epoch when the source became brighter and harder. In accreting pulsars, the CRSFs have been observed to vary or disappear with luminosity and spectral hardness \citep{vybornov2017,roy2025}. As the accretion rate increases, the height of the accretion column in a strongly magnetized neutron star system is predicted to increase \citep{lyub1988}. This could result in a change in the altitude of the line-forming region and therefore the local magnetic field strength, or it could result in a favorable geometric beaming pattern in which the CRSF becomes visible along our line-of-sight.

If we identify the 1.97 keV absorption feature as an electron CRSF, then the cyclotron line energy is given by the relation \citep{staubert2019}:
\begin{equation}
E_{\rm cyc,e} \approx 11.57 (1+z_g)^{-1} B_{12} \text{ keV},
\end{equation}
where $z_g$ is the gravitational redshift and $B_{12}$ is the magnetic field in units of $10^{12}$~G. The implied magnetic field strength is $B \approx 2.2 \times 10^{11}$~G (for $z_g=0.3$). We calculated a magnetospheric radius, $R_{\rm m} \approx 38.6$~km, for a typical NS ($M_{\rm NS}=1.4 M_{\odot}$ and $R_{\rm NS}=10^{6}$ cm) using Equation (22) in \cite{mushtu2015} for this field strength and the observed luminosity  ($L_{\rm X}\approx 1.13\times 10^{41}$~erg~s$^{-1}$).

While ultra-strong magnetic fields can allow extreme luminosities by heavily suppressing electron scattering cross-sections \citep{canuto1971,harding2006}, fields near $B \sim 10^{11}$~G result in cross-sections that remain close to the Thomson value. Consequently, the maximum luminosity for such a field strength is limited to approximately $10^{39}$~erg~s$^{-1}$ \citep{mushtu2015}. Since our luminosity exceeds this theoretical limit by two orders of magnitude, an electron-CRSF interpretation would require extreme geometrical beaming to remain physically consistent \citep{abarca2021}.

Moreover, electron CRSFs are usually detected at energies $>10$ keV with accompanying harmonics and are typically broad with $\sigma_{\rm line}/E_{\rm line}\sim 0.1-0.5$ \citep{staubert2019}. However, the observed line is significantly narrow ($\sigma_{\rm line}/E_{\rm line} \approx 0.04$) in the 2024 epoch and more consistent with proton CRSFs \citep{ibra2002}. Furthermore, higher harmonics are expected to be intrinsically weak (as is observed) or absent for proton CRSFs due to highly suppressed higher-order scattering cross sections \citep{harding2006,cruz2026}.

If the feature is interpreted as a proton CRSF, the energy relation is given by \citep{ho2001,ibra2002}:
\begin{equation}
E_{\rm cyc,p} \approx 0.63 (1+z_{g})^{-1} B_{14}  \text{ keV},
\end{equation}
where $B_{14}$ is the magnetic field strength in units of $10^{14}$~G. This would imply a magnetar-level field strength of $B \approx 4.1 \times 10^{14}$ G (for $z_g=0.3$). The corresponding magnetospheric radius is estimated to be $R_{\rm m} \approx 3100$~km.

For a typical NS and the observed luminosity, we calculate a spherization radius of $R_{\rm sph} \approx 6100$~km using Equation (34) of \cite{mushtu2015}. If one assumes a global magnetic field of  $B \approx 4.1 \times 10^{14}$ G, the super-Eddington wind would likely be starved (since $R_{\rm m}$ becomes comparable to $R_{\rm sph}$), making it difficult to explain the observed soft X-ray emission in the spectral continuum. 

We note that \cite{middle2019} ruled out a global magnetic field of $B \sim 10^{15}$~G for the CRSF detected in M51~ULX-8 \citep{bright2018}. A similar configuration may apply to NGC~3583~X-1. The magnetic field is likely dominated by a strong multipolar component ($B_{\rm mult} \sim 10^{14}$--$10^{15}$~G) confined near the neutron star surface  \citep{israel2017}, which gives rise to the observed CRSF. This leaves a typical ULX dipole field ($B_{\rm dip} \lesssim 10^{12}$~G) near the magnetosphere, resulting in a compact magnetospheric radius ($R_{\rm m} \lesssim 100$~km)  and a deep super-critical funnel (since $R_{\rm m} \ll R_{\rm sph}$) that produces the observed soft X-ray emission. 

Finally, the inferred surface magnetic field strength of $B \sim 4 \times 10^{14}$~G places NGC~3583~X-1 within an emerging population of extreme ULXs and transient HLXs powered by hyperaccreting neutron stars having magnetar-strength surface magnetic fields. For example, \cite{cruz2026} recently reported a candidate 3.3~keV p-CRSF in NGC~4656~ULX-1, implying a comparable magnetic field strength of $(6-7)\times 10^{14}$~G. Similarly, the 4.5~keV feature in M51~ULX-8 \citep{bright2018} requires a surface field of $B \sim 10^{15}$~G. Furthermore, the 8.56~keV absorption line detected in the transient HLX of NGC~4045 remains a p-CRSF candidate \citep{bright2022}, corresponding to an even more extreme surface field.

\subsection{Propeller Mechanism in Action?}
\label{sec:propeller}

The source exhibits extreme long-term luminosity variability spanning a factor of $\sim 45$, ranging from a peak unabsorbed X-ray luminosity (0.3--10 keV) of $L_{\rm X} \approx 3.4 \times 10^{41}$ erg s$^{-1}$ (\textit{Swift}/XRT ObsID: 00034153058) down to a low state with $L_{\rm X} \approx 7.4 \times 10^{39}$ erg s$^{-1}$, as observed with \textit{Chandra} (ObsID 19381). This deep low state is corroborated by multiple non-detections during the long-term \textit{Swift}/XRT monitoring where the source flux likely fell below the instrumental sensitivity. 

The dramatic variability in the source luminosity is reminiscent of the centrifugal inhibition of accretion in the ``propeller'' regime \citep{illarionov1975}. Notably, a similar luminosity drop of a factor of $\sim 40$ was reported for M82~X-2 \citep{tsygankov2016}. Targeted X-ray observations during the faint states are required to definitively confirm the propeller effect in this system.

\subsection{On the Absence of Coherent X-ray Pulsations}
\label{sec:nopulsations}

We note that coherent pulsations cannot be strictly ruled out during the current observations since the upper limits we have derived for the pulsed fraction are higher than the observed pulsed fractions in known PULXs such as NGC~1313~X-2 \citep{satya2019}, and those predicted from simulations in the presence of strong beaming \citep{mushtu2021}.

Physically, pulsations can be washed out by multiple scatterings in the optically thick accretion envelope \citep{mushtu2017}. Moreover, the neutron star spin axis is predicted to rapidly align with the axis of the inner accretion disk due to the action of strong torques in super-Eddington accretion flows \citep{king2020}. X-ray emission, in this scenario, would be reprocessed by the walls of the accretion funnel rather than viewed directly, resulting in the suppression of coherent pulsations. This may explain why some ULXs such as M51~ULX-8 \citep{bright2018} show CRSFs but do not exhibit pulsations to date. Therefore, it is also possible that NGC~3583~X-1 may belong to a population of non-pulsing neutron star ULXs.

Further deep observations are required to definitively confirm the presence or absence of transient pulsations in this system.

\section{Summary and Conclusions}
\label{sec:summary}

We have presented a detailed broadband X-ray study of the transient HLX, NGC~3583~X-1, using available \textit{XMM-Newton}, \textit{NuSTAR}, \textit{Chandra}, and \textit{Swift}/XRT data. We summarize the main results and conclusions as follows:
\begin{enumerate}
\setlength{\itemsep}{0pt}
    \item The broadband continuum of the source is consistent with the Hard Ultraluminous (HUL) state, with a  spectral energy cut-off measured at $\sim 5$--6 keV. This implies a near face-on viewing geometry, looking directly down the evacuated accretion funnel.
    \item The source shows extreme variability in the long-term light curve with luminosities usually in the $L_{\rm X} \sim 10^{40}$--$10^{41}$ erg s$^{-1}$ range, episodically dropping to $L_{\rm X} \sim 7.4 \times 10^{39}$ erg s$^{-1}$ or lower, reminiscent of the propeller effect in neutron stars.
    \item We detect a highly significant ($\approx 4\sigma$) absorption feature at $\sim 2$ keV, which is consistent across multiple continuum models. The absorption line parameters are well constrained, with a line energy centered at $E_{\rm line} \approx 1.97 \pm 0.04$ keV, a line width of $\sigma_{\rm line} \approx 74 \pm 40 $~eV, and an equivalent width of $EW_{\rm line} \approx -67^{+27}_{-28}$ eV (with model M3).
    \item An ultra-fast outflow (UFO) scenario was found to be less likely for the line's origin due to the absence of expected accompanying absorption features, the inferred viewing geometry, and the extreme outflow velocity required by our photoionization model ($v > 0.36c$).
    \item The line was interpreted as a candidate proton Cyclotron Resonance Scattering Feature (CRSF), implying that the source is a neutron star accretor with a local magnetic field strength of $B \sim 4 \times 10^{14}$~G. A global magnetar dipole field of this magnitude is disfavored as it would likely starve the super-Eddington wind responsible for the soft X-ray emission.
    \item The source likely exhibits a dipole magnetic field ($B_{\rm dip} \lesssim 10^{12}$ G) at larger distances up to the magnetosphere, and strong multipolar fields ($B_{\rm mult} \sim 10^{14}$ G) close to the neutron star surface.
\end{enumerate}

Put together, our results suggest that NGC~3583~X-1 likely belongs to a growing population of transient HLXs that host hyperaccreting, highly magnetized neutron stars, alongside sources such as NGC~5907~ULX1, and potentially NGC~470~HLX1 and NGC~4045~ULX. Deep broadband monitoring with current observatories, combined with upcoming missions (e.g., \textit{HEX-P}, \textit{NewAthena}), is essential to capture extreme propeller state transitions, characterize the CRSF, and search for coherent X-ray pulsations.

\section{Data Availability}

This research uses \textit{NuSTAR} archival data available at the High Energy Astrophysics Science Archive Research Center (HEASARC; \url{https://heasarc.gsfc.nasa.gov/cgi-bin/W3Browse/w3browse.pl}). The \textit{XMM-Newton} data are available in the \textit{XMM-Newton} Science Archive (XSA) at \url{https://www.cosmos.esa.int/web/xmm-newton/xsa}. This paper employs a \textit{Chandra} dataset, obtained by the \textit{Chandra} X-ray Observatory, contained in the Chandra Data Collection (CDC) `585'~\dataset[doi:10.25574/cdc.585]{https://doi.org/10.25574/cdc.585}. \textit{Swift}/XRT data products can be generated using the online automated pipeline tool \url{https://www.swift.ac.uk/user_objects/} provided by the UK Swift Science Data Centre (USSDC).

\begin{acknowledgments}

We thank the anonymous reviewer for their constructive suggestions and insightful comments that enhanced the quality of our work. The authors are grateful to Prof. Biswajit Paul for constructive suggestions that contributed to this work. KMJ is grateful to Dr. Manish Kumar and Mr. Akash Agarwal for their insightful inputs. KMJ is grateful to Dr. Radhakrishna V., GH-SAG, DD-PDMSA, and Director-URSC (ISRO), for their encouragement and continuous support to carry out this research.

This research has made use of data and/or software provided by the High Energy Astrophysics Science Archive Research Center (HEASARC), which is a service of the Astrophysics Science Division at NASA/GSFC. This research has made use of data obtained from the \textit{XMM-Newton} Science Archive, which is managed by the European Space Agency (ESA).  This research has made use of data from the \textit{NuSTAR} mission, a project led by the California Institute of Technology, managed by the Jet Propulsion Laboratory, and funded by the National Aeronautics and Space Administration. Data analysis was performed using the \textit{NuSTAR} Data Analysis Software (NuSTARDAS), jointly developed by the ASI Science Data Center (SSDC, Italy) and the California Institute of Technology (USA). 
This research has made use of data obtained from the \textit{Chandra} Data Archive provided by the \textit{Chandra} X-ray Center (CXC). This work made use of data supplied by the UK Swift Science Data Centre (UKSSDC) at the University of Leicester.
\end{acknowledgments}

\appendix

\section{Assessing Possible Contamination from the LLAGN in NGC 3583}
\label{app:agn}

NGC~3583~X-1 is situated at an angular distance of $\approx$1$\arcmin$ from the central low-luminosity AGN (LLAGN) of NGC~3583 \citep{baldi2018}. To check for potential contamination in our source spectrum, especially at the residual near 2 keV, we extracted and analyzed the LLAGN's spectrum using the 2024 \textit{XMM-Newton} data.

To model the LLAGN's spectrum we used the \texttt{TBabs*(mekal+powerlaw)} model describing emission from hot diffuse gas and a power-law continuum. We obtained a statistically acceptable fit (C-stat/d.o.f $\approx$ 1088.8/1182) with a power-law index of $\Gamma \approx 2.0$ and a plasma temperature of $kT_{e} \approx 0.52$~keV, likely associated with the diffuse gas in the immediate environment. The unabsorbed luminosity was estimated to be $L_{\rm X, LLAGN} \approx 1.3 \times 10^{40}\ \mathrm{erg\ s^{-1}}$  in the 0.3--10.0 keV band. The corresponding  unabsorbed flux was $F_{0.3-10,\rm  LLAGN} \approx 1.2 \times 10^{-13}\ \mathrm{erg\ s^{-1}\ cm^{-2}}$, which is $\sim 8$ times fainter than NGC~3583~X-1 ($F_{0.3-10,\rm src} \approx 9.6 \times 10^{-13}\ \mathrm{erg\ s^{-1}\ cm^{-2}}$). We concluded that the contamination from the LLAGN into our source extraction region was negligible due to the relatively large angular separation and the significantly lower flux.

We also added a \texttt{gabs} component to the LLAGN's spectrum fixing the line energy at 1.97 keV. However, the line width and strength could not be constrained, indicating the absence of a statistically significant feature. This confirms that the line detected in NGC~3583~X-1's spectrum is intrinsic to it and is not due to contamination from the nearby galactic nucleus.

\bibliography{citations}{}
\bibliographystyle{aasjournalv7}

\end{document}